\documentclass[amsmath,twocolumn,superscriptaddress,showpacs,aps,prb]{revtex4}
\usepackage{bm}
\usepackage{graphics}
\begin{document}

\title{Spin detection in quantum dots by electric currents}
\author{Eugene G. Mishchenko}
\affiliation{Lyman Laboratory of Physics, Harvard
University, MA 02138, USA}
\affiliation{L.D. Landau Institute for
Theoretical Physics, Moscow 117334, Russia}
\author{Arne Brataas}
\affiliation{Department of Physics, Norwegian University of
Science and Technology, N-7491 Trondheim, Norway}
\author{Yaroslav Tserkovnyak}
\affiliation{Lyman Laboratory of Physics, Harvard
University, MA 02138, USA}

\begin{abstract}
We develop a theoretical description of transport through quantum
dots connected  to reservoirs via spin-polarized ballistic
contacts. Rate equations account for spin accumulation inside the
dot, electron-electron interactions, and stochastic fluctuations.
It is shown that both the ac response (admittance) and the
frequency-dependent shot noise are governed by spin-flip
scattering, which can be used to detect spin polarization and
spin-flip processes in the dot.
\end{abstract}


% 73.63.Kv Electronic transport in quantum dots
% 72.25.-b Spin polarized transport
% 72.70.+m Noise processes and phenomena
% 75.47.De Giant magnetoresistance

\maketitle

The discovery of the giant-magnetoresistance effect initiated a
large activity aimed at understanding the transport properties of
nanoscale devices involving ferromagnetic elements
\cite{Gijs:adp97}. In such systems, transport is governed not only
by the charge flow, but also by the spin flow, spin precession,
and spin relaxation. The spin degree of freedom can provide new
functionality in electronic devices and is
envisaged as a future route of information processing. Until
recently, the experimental and theoretical efforts have been
mostly devoted to the exploration of the average current vs
voltage-bias characteristics at a given static magnetic
configuration. Recent fascinating developments concern
spin-current-induced magnetization dynamics \cite{SBT}. However,
time-dependent transport properties in circuits with static magnetic
elements have so far attracted little attention.

Current fluctuations are significant in submicron and nanoscale
devices and provide additional information about
correlation effects in such devices than the average conductance
\cite{Beenakker:pt03,BlB}. However, the information obtained from
the second moment of current fluctuations (noise power) is often
reduced by its universality, i.e., its independence of the
particular realization of the device. There are (at least) two
routes to achieve a further understanding beyond universality. The
first option is to look for characteristics beyond a simple noise
power, by elaborating the third (and higher) moments of current
fluctuations. The second way is to search for mechanisms that
break universality already for the second moment. In this Letter
we follow the second route and report a way to study
spin-dependent transport in quantum dots with spin-polarized
contacts by measuring current fluctuations. The electron spin
governs transport properties in spintronic and magnetoelectronic
circuits. We show that noise properties depend on and can be
used to study spin accumulation and spin-flip relaxation in
quantum dots.

Surprisingly few studies exist on the noise in ferromagnet
(\textit{F})/normal-metal (\textit{N}) systems. Ref.\
\onlinecite{Bulka:prb99} found super-Poissonian noise in
ferromagnetic single-electron transistors. Shot noise in the
\textit{F}/quantum-dot/\textit{F} system in the Coulomb-blockade
regime was also considered in Refs.\ \onlinecite{LL,SEJ,LS}. The
dependence of the noise on the relative magnetization orientation
of the leads have been computed in double-barrier systems with
tunnel, ballistic, or diffusive contacts \cite{Tserkovnyak:prb01}.
Effects of a spin-flip scattering on the shot noise have been
considered in Ref.\ \onlinecite{EM} using the Boltzmann-Langevin
approach for diffusive conductors connected to polarized leads.
Its generalization to multi-terminal systems has been given in
Ref.\ \onlinecite{BZ}.  Ref.\ \onlinecite{Lam} studied similar
effects using the Keldysh formalism assuming different chemical
potentials for spin-up and spin-down electrons in the leads. The
latter requires the system to be embedded into a larger
non-equilibrium \textit{F}/\textit{N} circuit which might have its
own noise characteristics, though. Shot noise of spin-polarized
entangled electrons has been considered in Ref.\ \onlinecite{EBL}
in quantum dots and wires. Effects of Andreev reflection have
recently been analyzed in Ref.\ \onlinecite{solicited} for an
F-N-S system.

We study dc conductance, ac admittance and noise through quantum
dots coupled via spin-polarized ballistic contacts. We go beyond the above-mentioned studies by computing the
frequency-dependent admittance and the frequency-dependent noise,
including electron-electron interactions. It is shown that the
spin degree of freedom reveals itself both in the ac and dc
properties. As we show below, the universality of the Fano factor
is relaxed if the attached leads are spin polarized.

Quantum dots are spatially-confined conducting islands Ohmically
contacted by metallic regions (reservoirs) and capacitively
coupled by electrostatic gates. The electron motion inside the dot
is usually chaotic as a result of the random scattering from
boundaries and/or impurities. The size of the dot $A$ enters its
spectral and transport properties through the mean level spacing
$\Delta=(\nu A)^{-1}$, which is inversely proportional to the
density of states at the Fermi level, $\nu$. Throughout our
discussion we assume the dot to be sufficiently large and open,
for the effects of the Coulomb blockade to be unimportant.
Nevertheless, Coulomb interactions are crucial and must be
included via the capacitance of the dot. We consider a system
where the central island of the dot is coupled to the left (L) and
right (R) reservoirs via ballistic point contacts (PC's)
supporting respectively $M_L$ and $M_R$ channels (per allowed spin
direction).

Let us first review key results for the nonmagnetic systems. The
theory for current fluctuations \cite{JPB,Naz,BS} is in a good
agreement with experiment \cite{O}. The average current
$\overline{J}= V/R$ at voltage $V$ is related to the resistance
$R= (h/2e^2) (M_{\text{L}}^{-1}+M_{\text{R}}^{-1})$. (We disregard
the quantum weak-localization corrections.) The fluctuations are
typically characterized by the current-current correlation
function
\begin{equation}
S(\omega)=\int dt~ e^{i\omega (t-t')}
\overline{\delta J(t) \delta J(t')} \, ,
\end{equation}
which for the dot at zero frequency and temperature is expressed
via the Fano factor $F={M_{\text{L}} M_{\text{R}}}/{M^2}$
($M=M_{\text{L}}+M_{\text{R}}$) as $S(0) = e \overline{J} F$. The
Fano factor, $F\le 1$, relates fluctuations at zero temperature
(i.e., the shot noise) to the average current and describes the
suppression of the fluctuations below the Poissonian value $F=1$
characteristic for independent electron transmission (as, e.g., in
tunnel junctions and vacuum diodes). For symmetric ($M_L=M_R$)
chaotic quantum dots, $F=1/4$, independent of its particular shape
and realization.

We now turn to the study of a quantum dot with spin-polarized
leads. We mainly focus on the most simple and illuminating case of
two fully spin-polarized leads. In addition we also consider the case where only one the leads is spin polarized. We first present our results, before outlining the derivation.

{\it Average dc current.} We compute the  average current  through
the quantum dot and find the resistance in the antiparallel
configuration,
\begin{equation}
R_{\text{a}}=R_{\text{L}}+R_{\text{sf}}+R_{\text{R}} \, ,
\label{Ra}
\end{equation}
%,
where the resistances of the contacts are $R_{\text{L,R}} =
h/(e^2M_{\text{L,R}})$ and the characteristic spin-flip resistance
is $R_{\text{sf}} = 2\Delta\tau_s/e^2$, in terms of the spin-flip
relaxation time $\tau_s$ \cite{Brataas:prb99}. The resistance can
be understood in terms of a circuit consisting of a series of the
spin-polarized left and right PC resistances and a spin-flip
resistance coupling two PC's \cite{Brataas:prb99}. Similarly, in
the parallel configuration,
\begin{equation}
R_{\text{p}}=R_{\text{L}}+R_{\text{R}} \, .
\label{Rp}
\end{equation}
The magnetoresistance of the dot is therefore determined by the
spin-flip scattering rate: $R_{\text{ap}}-R_{\text{p}}=R_{\text{sf}}$.

{\it Admittance.} General expressions for the frequency dependent
admittance and noise in the arbitrary asymmetric geometry are
lengthy, so we present below only the results for the symmetric
($M_{\text{L}} = M_{\text{R}}=M/2$) dot. The admittance is a
$2\times2$ matrix, which for a symmetric dot is determined by just
two quantities.
 Consider ac bias applied to the
left lead, $V(t)=V_\omega e^{-i\omega t}$, while the right lead is
kept at equilibrium. The current response $J_L(\omega)
=G^{LL}(\omega) V_\omega$, $J_R(\omega) =G^{LR}(\omega) V_\omega$
is determined by the left, $G^{LL}$, and right, $G^{LR}$,
admittance. In the parallel configuration the ac response is independent of the
spin-flip relaxation,
\begin{eqnarray}
G^{LL}_{\text{p}}=\frac{e^2M}{4h} \frac{1 -2i\omega\tau_C}
{1-i\omega\tau_C},~ %\,,\nonumber\\
G_{\text{p}}^{LR}=\frac{e^2M}{4h} \frac{1} {1-i\omega\tau_C}\, ,
\label{adm_p}
\end{eqnarray}
where $\tau_C$ is the electrochemical $RC$ time of the dot
\begin{equation}
\label{admitt}
\frac{1}{\tau_C}=\frac{1}{\tau_d}+\frac{e^2M}{hC}\,,
\end{equation}
$C$ is the capacitance of the dot, and
$\tau_d=h/(\Delta M)$ is the electron dwell time. In the
antiparallel configuration,
\begin{equation}
\left\{ \begin{array}{c} G^{LL}_{\text{a}}\\
G^{LR}_{\text{a}}\end{array} \right\} =
\left\{ \begin{array}{c} G^{LL}_{\text{p}}\\
G^{LR}_{\text{p}}\end{array} \right\} \frac{\gamma-
2i\omega\tau_d} {1+\gamma- 2 i\omega\tau_d} \label{adm_a}
\end{equation}
with  $\gamma=2\tau_d/\tau_s$ being the normalized spin-flip rate.
For large spin-flip rates, $\gamma \to \infty$, equations
(\ref{adm_p}) and (\ref{adm_a}) coincide and reproduce the result
of Ref.\ \onlinecite{BB}. The admittance exhibits frequency
dependence on the scale of $~\tau_d^{-1}$, which in a typical
experimental situation in GaAs quantum dots is much smaller than
the charge relaxation rate, $\tau_d^{-1} \ll \tau_C^{-1}$. In the
limit, $\omega \to 0$, the admittance (\ref{adm_p}) and
(\ref{adm_a}) reduce to the static conductance,
 $G^{LL} \to G^{LR} \to R^{-1}$, as given by Eqs.\
 (\ref{Ra}-\ref{Rp}).

{\it Current fluctuations}.  The ac response reveals strong
dependence on the polarization of the leads and the spin-flip
relaxation time, being governed by the degree of spin accumulation
in the dot. Similar effects are revealed in the magnitude and
spectrum of current fluctuations accompanying the dc current.
Since the noise of a mesoscopic scatterer depends on its
energy-relaxation processes, we make in the following a
distinction between the elastic and inelastic transport regimes.
We will restrict our attention to noise at small frequencies,
$\hbar\omega\ll\max(eV,k_BT)$, and assume a typical situation when
the charging energy exceeds the mean level spacing,
$e^2/C\gg\Delta$.

{\it Elastic transport through the dot.} First we compute the Fano
factor in the parallel configuration,
\begin{eqnarray}
\label{Fp_el}
F_{\text{p}}^{\text{e}}(\omega)=F_\omega+\frac{2(F_\omega-\frac{1}{4})}
{\gamma (1+\omega^2\tau_s^2)},~~ F_\omega = \frac{1+2\omega^2
\tau_C^2}{4(1+\omega^2 \tau_C^2)} \, . \label{Fp}
\end{eqnarray}
which in the typical scenario $\tau_d,\tau_s \gg \tau_C$ is very
weakly dependent on the spin-flip scattering. Both for strong
$\tau_s \to 0$ and weak $\tau_s \to \infty$ spin-flip scattering
the expression (\ref{Fp_el}) is reduced to the Fano factor for the
unpolarized dot $F_\omega$ exhibiting a crossover from the
universal value of $F=1/4$ at low frequencies, $\omega \ll
\tau_C^{-1}$, to $F=1/2$ at high frequencies, $\omega \gg
\tau_C^{-1}$. To the best of our knowledge the expression for
$F_\omega$ has never been explicitly presented before. [A general
formalism has recently been presented for the noise in
spin-independent mesoscopic conductors in Ref.\ \onlinecite{Nag}
that, upon further elaboration, reproduces $F_\omega$].

The Fano factor in the antiparallel configuration differs
significantly from $F_\omega$,
\begin{equation}
\label{Fa_el} F_{\text{a}}^{\text{e}}(\omega) =
\frac{F_\omega(2+\gamma)\gamma^2(1+\omega^2\tau_s^2)+1+\gamma+\gamma^2/2}
{(1+\gamma)[(1+\gamma)^2+\gamma^2\omega^2\tau_s^2]},
\end{equation}
In the static
regime, $\omega=0$, charging effects are irrelevant and the Fano
factor reduces to
\begin{equation}
F_{\text{a}}^{\text{e}}(0)=\frac{1+\gamma+\gamma^2+\gamma^3/4}{(1+\gamma)^3}.
\label{Fa_el_stat}
\end{equation}
This formula describes a smooth crossover between Poissonian noise
$F=1$ at $\gamma\to 0$ and a conventional value $F=1/4$ at $\gamma
\to \infty$ characteristic for a spin-independent shot noise. The
full Poissonian noise for weak spin-flip scattering is reminiscent
of the shot noise in tunnel junctions. Indeed, current is
suppressed at $\gamma\to 0$ and the transport through the dot is
mostly due to the electrons that flip spin only once. Those
electrons are few and therefore move independently inside the dot.
Around $\omega \sim \tau_s^{-1}$ the Fano factor changes from the
static value (\ref{Fa_el_stat}) to $(1+\gamma/2)/[2(1+\gamma)]$
for $\tau_s^{-1} \ll \omega \ll \tau_C^{-1}$. With further
increase of the frequency, the Fano factor rises up to
$(1+\gamma/2)/(1+\gamma)$ for $\omega \gg \tau_C^{-1}$.
Experimentally, a situation with only one of the leads polarized
might be of interest. The Fano factor for the zero-frequency noise
in the elastic regime changes from $F= 1/4$ for small spin-flip,
$\gamma \to 0$, to a slightly smaller value $F=2/9$ for large
spin-flip, $\gamma \to \infty$.

The noise at finite temperatures in the elastic regime is
calculated as well and expressed via the corresponding Fano
factors in a usual way \cite{formula}. The situation is more
interesting in the inelastic regime.

{\it Inelastic transport through the dot}. At elevated
temperatures the (spin-conserving) electron-phonon scattering
thermalizes the electronic distribution function. In the antiparallel
configuration, the finite spin-flip relaxation prevents
equilibration of electrons with opposite spin polarizations. The
noise becomes
\begin{equation}
\label{Fa_in}
\frac{S_{\text{a}}^{\text{i}}}{e\overline{J}}=\frac{2k_BT\gamma}{eV(1+\gamma)}
+\frac{1}{(1+\gamma)^2}\coth{\left[\frac{eV}{2k_BT
(1+\gamma)}\right]}\,.
\end{equation}
The second term in Eq.\ (\ref{Fa_in}) is the nonequilibrium
contribution due to the spin-flip scattering in the antiparallel
configuration. The parallel configuration of the leads reveals the
Nyquist-Johnson thermal noise only:
\begin{equation}
\label{Fp_in}
S_{\text{p}}^{\text{i}}/(e\overline{J})=2k_BT/eV.
\end{equation}
which vanishes in the low-temperature limit, $k_B T \ll eV$.

{\it Derivation.} Let us now outline the derivation of our
results. The spin dynamics in the dot is determined by the rate
equations for the deviations of the numbers of electrons
$N_\uparrow$ and $N_\downarrow$ from their equilibrium values
[assuming $\hbar\omega\ll\max(eV,k_BT)$]:
\begin{eqnarray}
\label{rate} \frac{\partial N_\uparrow}{\partial t}= J_L/e -
 \frac{N_\uparrow-N_\downarrow}{2\tau_s}  +{\cal L}(t),\nonumber\\
\frac{\partial N_\downarrow}{\partial t}= J_R/e +
\frac{N_\uparrow-N_\downarrow}{2\tau_s} -{\cal L}(t)\,,
\end{eqnarray}
here $J_L$ and $J_R$ are the electric currents entering the dot
through the left and right PC's, respectively, and
${\cal L}(t)$ is a Langevin source due to the randomness of the
spin-flip events. The current through the left quantum PC is
\begin{equation}
\label{current} \frac{J_L}{e}= \frac{M_L }{h}\left( \int d\epsilon
[n_L - n_0]- {N_\uparrow}\Delta- \frac{eQ}{C}\right)+{\cal
I}_L(t)\,,
\end{equation}
where  $Q$ denotes the charge imbalance in the dot: $\partial Q/
\partial t =J_L+J_R$. A similar  relation exists for the current through
the right contact. The first two terms in the parentheses of Eq.\
(\ref{current}) represent simply the current through the ballistic
constriction, $n_L$ is the distribution function in the left lead
and $n_0(\epsilon)$ is its equilibrium value. The third term
describes the charging effect. Electrons experience the charge
imbalance $Q$ in the dot as an effective change of the chemical
potential $eQ/C$ (same for spin-up and spin-down electrons) of the
central island having a capacitance $C$. The last term ${\cal
I}_L(t)$ is the Langevin contribution describing intrinsic
stochastic noise in the PC. This term vanishes on average and has
the correlator with the white-noise spectrum \cite{BlB}:
\begin{eqnarray}
\label{corrsp} \overline{{\cal I}_L(t){\cal I}_L(t')}&=&\delta
(t-t') \frac{M_L}{h} \nonumber\\ && \times \int d\epsilon [n_L
(1-n_L) +\overline{f}_\uparrow(1-\overline{f}_\uparrow)],
\end{eqnarray}
and similarly for the right quantum PC, but with $\overline{f}_\downarrow$
on the right-hand side. The mean (averaged over fluctuations)
distribution of electrons inside the dot, $\overline{f}_\uparrow
(\epsilon)$, $\overline{f}_\downarrow (\epsilon)$, is homogeneous
and isotropic (as a result of the multiple chaotic scattering).
The Langevin term ${\cal L}(t)$ in Eq.\ (\ref{rate}) describes the
randomness of the spin-flip relaxation \cite{EM}:
\begin{eqnarray}
\label{spinflip} \overline{{\cal L}(t){\cal L}(t')}= \frac{\delta
(t-t')}{2\Delta\tau_s} \int d\epsilon
[\overline{f_\uparrow}(1-\overline{f}_\downarrow)+\overline{f}_\downarrow(1-\overline{f}_\uparrow)]\,.
\end{eqnarray}
This expression implies that separate events of  spin-flip
relaxation are independent and therefore Poissonian. They are also
assumed to be elastic and have a constant scattering rate around
the Fermi energy.

{\it Elastic transport through the dot.} When the inelastic (e.g.,
electron-phonon) scattering time exceeds the dwell time in the
dot, the energy of an electron is conserved during its transport
across the dot. The mean electron distribution function is then
found from the condition that the incoming flow of spin-up
electrons through the left PC equals the total rate of spin-flip
transition (since there are no spin-up electrons leaving the dot
via the right quantum PC),
\begin{eqnarray}
\frac{M_L\Delta}{h}\left[n_L(\epsilon) -
\overline{f}_\uparrow(\epsilon)\right]=
\frac{\overline{f}_\uparrow(\epsilon)-\overline{f}_\downarrow(\epsilon)}{2\tau_s}\,.
\end{eqnarray}
with a similar condition holding for spin-down electrons in the
right PC. Combining the two relations, we obtain (omitting the
energy dependence)
\begin{eqnarray}
\label{distribution}
\left\{\begin{array}{c} \overline{f}_\uparrow\\
\overline{f}_\downarrow \end{array} \right\} =\frac{ \left\{\begin{array}{c} n_L\\
n_R \end{array} \right\}+\gamma\overline{f} }{1+\gamma}\,,
~~~\overline{f}= \frac{n_LM_L+n_RM_R}{M_L+M_R}\,.
\end{eqnarray}
For large spin-flip scattering rate, $\gamma \gg 1$, the distribution
function becomes spin independent and equal to $\overline{f}$.

Let us calculate the average current through the dot under dc
bias, $n_L(\epsilon)=n_0(\epsilon-eV)$,
$n_R(\epsilon)=n_0(\epsilon)$. In a steady state the dot is
electrically neutral, $Q=0$. Substituting $\overline{N}_\uparrow
=\Delta^{-1}\int
d\epsilon[\overline{f}_\uparrow(\epsilon)-n_0(\epsilon)]$, we
obtain the current through the dot $\overline{J}=V/R$ with the
resistance in the parallel and antiparallel configurations given
by equations\ (\ref{Ra}) and (\ref{Rp}) respectively.

The stochastic Langevin sources ${\cal I}$ and ${\cal L}$ describe
the fluctuations around the mean solution, Eq.\
(\ref{distribution}). The charge imbalance is given by $Q=e(\delta
N_\uparrow + \delta N_\downarrow)$, with the fluctuating
quantities $\delta N_\uparrow$, $\delta N_\downarrow$ determined
from the solution of the coupled inhomogeneous equations
(\ref{rate}) around the steady state. The fluctuation of the
electric current is then found from Eq.\ (\ref{current}). The
calculations are rather straightforward in the Fourier
representation. E.g.\ in the case of a symmetric dot ($M_L=M_R$)
with charging energy exceeding the mean level spacing ($e^2/C \gg
\Delta$) and the antiparallel orientation of the leads, we find,
\begin{equation}
\label{deltaj} \frac{\delta
J_L}{e}=\frac{\gamma(1-i\omega\tau_s)[(\frac{1}{2}-i\omega\tau_C){\cal
I}_L-{\cal I}_R]-(1-i\omega\tau_C){\cal
L}}{(1-i\omega\tau_C)(1+\gamma-i\gamma \omega\tau_s)}\,.
\end{equation}
The calculation of the noise power is now performed with the help
of the correlators (\ref{corrsp}) and (\ref{spinflip}) giving
the frequency-dependent Fano factors in (\ref{Fp_el}) and (\ref{Fa_el}).

{\it Inelastic transport through the dot.} When the temperature
$T$ increases, the inelastic scattering changes dramatically the
noise pattern. Once the characteristic electron-phonon scattering
time drops below the dwell time in the dot, the electron
distribution $\overline{f}_\alpha(\epsilon)$ relaxes to the
Fermi-Dirac distribution. The chemical potentials of spin-up
 and spin-down  electrons are found from the
current-conservation conditions for the right and left PC's. For
the antiparallel configuration: $\mu_\uparrow=eV(M+\gamma
M_L)/[(1+\gamma)M]$, $\mu_\downarrow=eV\gamma M_L/[(1+\gamma)M]$.
%\begin{equation}
%\mu_\alpha=eV \frac{\delta_{\alpha \uparrow} (M_{L}+M_R) +\gamma
%M_L}{(1+\gamma)(M_L+M_R)}\,,
%\end{equation}
 It can be easily verified
that the resistance  in the inelastic regime is still given by
Eq.\ (\ref{Ra}).  The above expression (\ref{deltaj}) for the
fluctuation of electric current holds in this case as well.
Calculating the static noise power in a symmetric dot at finite
temperature with the help of correlators\ (\ref{corrsp}) and
(\ref{spinflip}), we find expressions (\ref{Fa_in}) and
(\ref{Fp_in}).

{\it In conclusion}, we have presented a framework for the
computation of spin-dependent transport in quantum  dots based on
the Langevin rate equations. Its application is illustrated by
finding the ac admittance and the frequency-dependent noise in
dots coupled with reservoirs via spin-polarized ballistic
contacts. Time-dependent measurements seem to be a much more
powerful tool for studying spintronic effects than dc
measurements. In particular, the knowledge of the dc
magnetoresistance determines only the ratio of the spin-flip
scattering rate $\tau^{-1}_s$ to the mean level spacing $\Delta$.
Moreover, when $\tau^{-1}_s \gg M\Delta$ the magnetoresistance
represents only a small fraction of the total resistance, making
the former difficult to extract. On the other hand, time-dependent
measurements of the ac admittance and noise spectrum should reveal
$\tau_s$ without any restrictions imposed by the magnitude of the
mean level spacing and the number of channels coupled to the dot.

We are grateful to B.\ I.\ Halperin for useful discussions and his
great hospitality during A.B.'s\ stay at Harvard University where
most of this work was carried out. Communications with Ya.\ M.\
Blanter are also acknowledged. The work was supported by the NSF
Grant PHY-01-17795, DARPA Award No. MDA 972-01-1-0024, and by the
Harvard Society of Fellows.

\end{document}